\documentclass[a4paper, twoside, final, runningheads, twocolumn]{raa_twocolumn}           

\newcommand{\sech}{sech}
\usepackage{graphicx}
\usepackage{amssymb,amsmath}
\usepackage{txfonts}
\usepackage{natbib}
\usepackage{threeparttable, tablefootnote}
\bibpunct{(}{)}{;}{a}{}{,} 
\usepackage[colorlinks = true, citecolor = blue]{hyperref}
\renewcommand*{\tablefootnote}\textsuperscript{{\alph{footnote}}}

\begin{document} 

   \title{Milky Way globular cluster dynamics: are they preferentially co-rotating?}
   
   \volnopage{ {\bf 20XX} Vol.\ {\bf X} No. {\bf XX}, 000--000}
   \setcounter{page}{1}

   \author{Saikat Das\inst{1}
          \and
          Nirupam Roy\inst{2}}
          \email{saikatdas@rri.res.in}
   \institute{Astronomy \& Astrophysics Group, Raman Research Institute, Bangalore 560080, India
          \and
          Department of Physics, Indian Institute of Science, Bangalore 560012, India}
 
  \abstract{The motion of the baryonic components of the Milky Way is governed by both luminous and dark matter content of the Galaxy. Thus, the dynamics of the Milky Way globular clusters can be used as tracers to infer the mass model of the Galaxy up to a large radius. In this work, we use the directly observable line-of-sight velocities to test if the dynamics of the globular cluster population is consistent with an assumed axisymmetric gravitational potential of the Milky Way. For this, we numerically compute the phase space distribution of the globular cluster population where the orbits are either oriented randomly or co-/counter- rotating with respect to the stellar disk. Then we compare the observed position and line-of-sight velocity distribution of $\sim$ 150 globular clusters with that of the models. We found that, for the adopted mass model, the co-rotating scenario is the favored model based on various statistical tests. We do the analysis with and without the GCs associated to the progenitors of early merger events. This analysis can be extended in the near future to include precise and copious data to better constrain the Galactic potential up to a large radius.
}
   
\titlerunning{Milky Way globular cluster dynamics}
\authorrunning{Das \& Roy}
   \maketitle

\section{Introduction}

The nearly flat rotation curve of the Milky Way (MW) at outer Galaxy, inferred from the stellar motion as well as the spectroscopic observation of a variety of tracers of the interstellar medium (e.g., H$\alpha$, H{\sc i} and CO emission lines), is explained by invoking the existence of a massive dark matter halo \citep[e.g.,][]{Rubin80, Begeman91}. Although there are a few galaxies for which the rotation curve falls off according to a Keplerian prediction \citep{Honma97}, the majority of spiral galaxies exhibit a similar flat rotation curve. The nature and properties of this dominant component of the mass in our Galaxy, at present, remain mostly uncertain. The mass and density distribution of various components of the Milky Way have been studied earlier in detail through mass models \citep[e.g.,][]{Caldwell81,DB98,Klypin02}, kinematic models \citep[e.g.,][]{Sharma11} and dynamical models \citep[e.g.,][]{Widrow08, Bovy_2013}. The most notable one among these is the mass model put forward by \citet{DB98}, which considers an axisymmetric potential and three principal components of the MW, viz. the disk, the bulge, and the halo. The disk consists of the interstellar medium (ISM), and the thin and thick stellar disks. The bulge and the halo are each described by a spheroidal density distribution. The H{\sc i} 21-cm line, in particular, is one of the most powerful tools to study the kinematics of our Galaxy, as the radial extent of the H{\sc i} gas is greater than that of the visible component. However, the dynamics of the Galaxy and, in turn, the properties of the dark matter halo can also be studied from the structure and kinematics of other baryonic components, such as the globular clusters and the satellite galaxies. In this work, we use the phase space distribution of the globular clusters based on direct observables (position and line-of-sight component of their velocity) to check the consistency of the current Milky Way mass model.

The hierarchical structure formation predicts that the merging of smaller 
subhaloes leads to the formation of dark matter halo \citep[e.g.,]{Wang08, 
Frenk12}. The residual subhaloes are observed today as satellite galaxies. The 
Milky Way has about 59 satellite galaxies (SGs) within 0.5 Mpc from the 
Galactic center that are gravitationally bound to the Milky Way, but not all 
are necessarily in orbit \citep{Kallivayalil06, Besla07, Pardy19}. The Milky 
Way also has nearly 200 globular clusters (GCs) with a roughly spheroidal 
distribution around the Galaxy. They constitute the halo population of our 
Galaxy. Majority of the GCs lie at low latitudes in the inner Galaxy, i.e., 
within $\sim$ 20 kpc from the Galactic center \citep{Koposov07, Dotter11, 
Gaia18}. The satellite galaxies are more dark matter dominated than the 
globular clusters in their small-sized halos. In the outermost regions of the 
Galaxy, beyond the luminous disk, the gravitational potential of the dark 
matter halo can be constrained from observed velocities of GCs and SGs 
\citep{Sofue13}.

Depending on the nature of dark matter (DM) candidate, the number of subhaloes predicted from cosmological simulations can be as much as few orders of magnitude more than the number of dwarf galaxies observed as satellites \citep{Strigari08, McConnachie09, Strigari18}. The general consensus in the $\Lambda-$CDM model is that the stellar halos of MW type galaxies are formed from continuous accretion, merger events, or tidal disruption of many smaller DM subhalos at high redshifts \citep{Bullock01}. The outer- and inner-halo of the Milky Way with overlapping structural components can thus be assumed to be composed of two kinds of stellar populations. One that originated in other galaxies and was accreted by MW in merger events, and another which originated \textit{in situ} from the evolution of MW itself. They exhibit different spatial density profiles, stellar orbits, and stellar metallicities \citep{Carollo07}. Chemodynamical studies of the latest data from {\it Gaia} \citep{Gaia16} and SDSS \citep{Abolfathi18} provides definitive evidence of the presence of tidal debris from a major merger event around 8-10 Gyr ago, during the early stages of halo assembly, leaving its imprint on the `sausage' like structure formed in the velocity-space \citep{Helmi18, Belokurov18, Myeong18}. Apart from Gaia-sausage, there are predictions of accretion due to other less-massive merging dwarfs \citep{Myeong19, Piatti19}.

In this paper, we investigate whether these accretion events can have contributed significantly to the resulting dynamics of the GC population. With limited information about the orbits, studying the exact dynamics of individual globular clusters or satellite galaxies is an intricate problem. Instead, we address here the consistency of the globular cluster phase space distribution for an assumed gravitational potential. We consider the dataset of GCs 
\citep{Harris96, Harris10, Sohn18} with known Galactic coordinates ($l, b$), distance to the clusters from the Sun, $D_\odot$, and the observed line-of-sight velocity $v_{\rm los}$. From this, we construct the distribution of GCs around the Galactic center. We then use the public-licensed code \textsc{GalPot} to simulate, for a given mass model of the Milky Way, the position-velocity ($l$ vs. $v_{\rm los}$) distribution for a sample of GCs with the same Galactrocentric distribution. A comparison between the simulated and the observed phase space distribution will allow one to check if the assumed mass model is consistent with the GC dynamics. We also check how the phase space distribution changes because of the GCs associated with the progenitor galaxies of the merger events -- Gaia-Enceladus, Sequoia, or Sagittarius dSph. Please note that the same analysis can be done for the satellite galaxies as well. 
However, the line-of-sight velocity data is available for a lesser number of SGs \citep{Newton17}, and hence, here we restrict our analysis mostly to GCs.

We discuss the observed statistics for GCs and the mass model used for this analysis in Sec.~\ref{sec:d&a}. The methodology of our analysis and the results are presented in Sec.~\ref{sec:results}. We discuss the implications of our results in Sec.~\ref{sec:disc} and draw our conclusions in Sec. \ref{sec:conclu}.

\section{Data and Mass Model} \label{sec:d&a}

The catalogue of the GCs \citep{Harris96, Harris10} provides the coordinates 
($l$, $b$), the line-of-sight velocity $v_{\rm los}$, the metallicity, 
photometry and other structural parameters. In the following, we discuss the 
observed statistics of the GCs and their spatial distribution in 
Subsec.~\ref{subsec:obs}, and present the potential model used to calculate the 
circular velocity from the Galactocentric distance $\mathbf{R_{G}}$ and the 
setup for \textsc{GalPot} code in Subsec.~\ref{subsec:grav_pot}. 

\subsection{Observed Statistics} \label{subsec:obs}

\begin{figure}[t]
\centering
	\includegraphics[trim={0 1.5cm 0 1.5cm}, clip, width=0.49\textwidth]{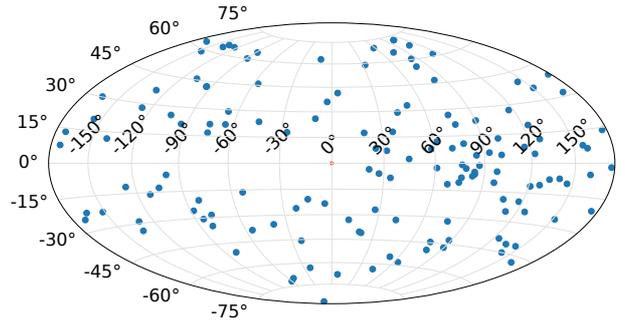}
    \caption{\small{Hammer projection of the Galactocentric distribution of globular clusters}}
    \label{fig:glob_ham}
\end{figure}

\begin{figure}
	\centering
	\includegraphics[width=0.49\textwidth]{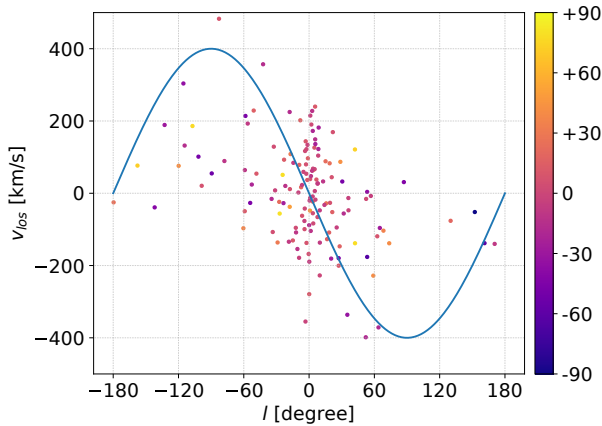}
    \caption{\small{Distribution of the line-of-sight velocity ($v_{\rm los}$) as a function of Galactic longitude ($l$) for globular clusters with colorbar indicating the Galactic latitude ($b$).}}
    \label{fig:glob_lv}
\end{figure}

For each GC, we calculate the Galactocentric distance $R_G$, vertical distance $z$ from the Galactic plane, and the Galactocentric angular coordinates $\theta$, $\phi$ from $l, b, D_\odot$ data of the GCs. For this, we took the distance to the Sun from the Galactic center to be $R_0=8.2$ kpc, the best fit value found in \cite{McMillan17}. By matching the best dynamical model obtained in \cite{Chatzopoulos15} to the proper motion and line-of-sight velocity dispersion data of nuclear star clusters, they have found the value of $R_0$ to be $8.27\pm0.09|_{\rm stat}\pm0.4|_{\rm syst}$, where the systematic errors account for uncertainties in the dynamical modeling. The number of GCs falls off sharply beyond $R_G=20$ kpc. The Galactocentric angular distribution of the GCs is shown using the Hammer projection in Fig.~\ref{fig:glob_ham}. As expected, the distribution is consistent with a uniform spherical distribution. We show the observed line-of-sight velocities $v_{\rm los}$ against Galactic longitude $l$ in Fig.~\ref{fig:glob_lv}, where the color bar indicates the value of the Galactic latitude $b$ for the GCs. The overplotted sinusoidal curve in the figure illustrates the component of $v_{\rm los} \propto \sin(l)$, reminiscent of similar boundary for tracers from the Galactic disk. Note that the mentioned catalog lists 157 sources, of which the velocity information is available for 143 GCs, and only those are used in the analyses.

\subsection{The gravitational potential} \label{subsec:grav_pot}

The observed distribution of ($l$, $v_{\rm los}$) of the GCs are compared with 
the predicted distribution for the model axisymmetric potential of the Milky 
Way. For this, we have used the public licensed code \textsc{GalPot}. It was 
originally written in C++ by Walter Dehnen and later developed by Paul J. 
McMillan \citep{McMillan17}, which is an extension of a previous model by 
\cite{McMillan11}. In addition to other components, the new model incorporates 
gas discs that account for Milky Way's cold gas. The MW mass is decomposed into 
6 axisymmetric components -- bulge; dark-matter halo; thin and thick stellar 
discs; and H{\sc i} and molecular gas discs. With an axisymmetric approximation 
to \cite{Bissantz02} model, the bulge density profile is given by
\begin{equation}
\rho_{\rm b} = \frac{\rho_{\rm 0,b}}{(1+r'/r_{\rm 0})^{\alpha}} \ \exp[-(r'/r_{\rm cut})^2] \,,
\end{equation}
where $r'=\sqrt{R^2 + (z/q)^2}$ in cylindrical coordinates, with $\alpha=1.8$, 
$r_{\rm 0} = 0.075$ kpc, $r_{\rm cut} = 2.1$ kpc and axis ratio $q=0.5$. The 
total bulge mass considered is $M_{\rm b} =8.9\times10^9$ $M_{\odot}$ with an 
uncertainty of $\pm$ 10\%. The scale density $\rho_{\rm 0,b}=9.93\times 
10^{10}$ $M_{\odot}$ kpc$^{-3}$ $\pm$ 10\%. The thin and thick stellar disc of 
Milky Way are modelled as exponential according to \cite{Gilmore83} model
\begin{equation}
\rho_{\rm d}(r, z) = \frac{\Sigma_{\rm 0}}{2 z_{\rm d}} \ \exp\bigg(-\frac{\mid z\mid}{z_{\rm d}} - \frac{R}{R_{\rm d}}\bigg) \,,
\end{equation}
with scale height $z_{\rm d}$, scale length $R_{\rm d}$ and central surface 
density $\Sigma_{\rm 0}$. The total disc mass is $M_{\rm d} = 2\pi \Sigma_{\rm 
0}R_{\rm d}^2$. The scale heights of the discs are fixed at $z_{\rm d,thin} = 
300$ kpc and $z_{\rm d,thick} = 900$ kpc. The H{\sc i} and molecular gas discs 
are given by the functional form mentioned in \cite{DB98} as given below
\begin{equation}
\rho_{\rm d}(R, z) = \frac{\Sigma_{\rm 0}}{4z_{\rm d}} \ \exp\bigg(-\frac{R_{\rm m}}{R} - \frac{R}{R_{\rm d}}\bigg)\sech^2(z/2z_{\rm d}) \,.
\end{equation}

Similar to stellar disc, the gas disc also exhibits an exponential decline with 
$R$, but has a hole in the center with an associated scale length $R_{\rm d}$. The 
maximum surface density is found at $R=\sqrt{R_{\rm d}R_{\rm m}}$, and the 
total disc mass is given by $M_{\rm d}=2\pi\Sigma_{\rm 0}R_{\rm d}R_{\rm m} 
K_{2}(2\sqrt{R_{\rm m}/R_{\rm d}})$ where $K_2$ is a modified Bessel function. 
Also, the disc model posses an isothermal $\sech$-squared profile. The H{\sc i} 
disc model resembles the distribution found in \cite{Kalberla08}. The presence 
of the gas discs significantly deepens the potential well near the Sun, and 
hence affects the dynamics near the solar neighbourhood. The surface density is 
set to be 10 $M_\odot$pc$^{-2}$ at a fiducial value of $R_{\rm 0} = 8.33$ kpc, 
the distance of Sun from the Galactic center. The dark matter halo density is 
described by
\begin{equation}
\rho_{\rm h} = \frac{\rho_{\rm 0,h}}{x^{\gamma}(1+x)^{3-\gamma}} \,,
\end{equation}
where $x={r/r_{\rm h}}$, with $r_{\rm h}$ the scale radius. They have 
considered $\gamma=1$, for the best-fit potential model, which is the NFW 
profile \citep{Navarro96}. \textsc{GalPot} provides the gravitational potential 
associated with axisymmetric density distributions. It includes the potential 
models from \cite{Piffl14}, \cite{McMillan11}, \cite{DB98}, \cite{McMillan17}, 
and their variants. Here, we use the best-fit potential model of 
\cite{McMillan17}, which contains 4 disk components and 2 spheroid components. 
The values of various parameters for this best-fitting potential model is given 
in Table 3 of \cite{McMillan17}. The mass of the Milky Way within 300 kpc, 
calculated by \cite{Watkins12} is found to be betweeen 1.2 and 2.7$\times 
10^{12}$ $M_{\odot}$. The estimate found by using \textsc{GalPot} is found to 
be $(1.6\pm0.3)\times10^{12}$ $M_\odot$, which is well within the plausible 
range. Thus, the potential model is representative and well-suited while 
considering the dynamics of GCs too.

\begin{figure*}
	\includegraphics[width=\textwidth]{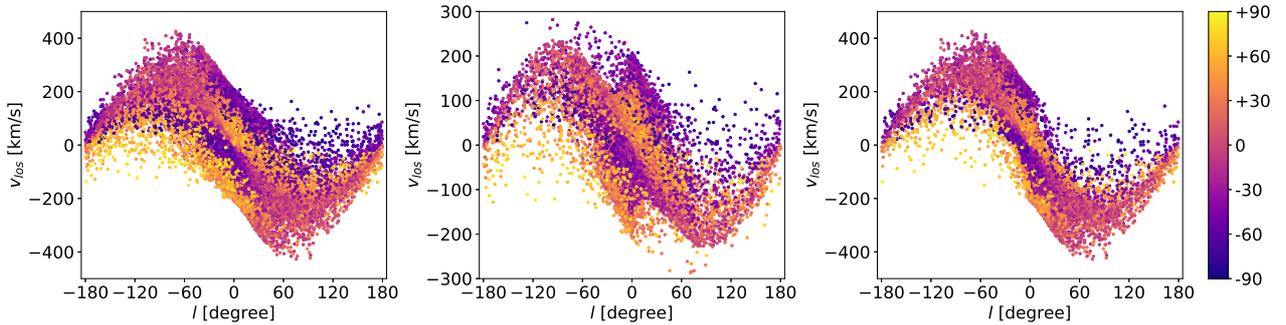}
    \caption{\small{$l$ vs. $v_{\rm los}$ plot for (a) {\it{Left:}} mixed rotation, (b) {\it{Middle:}} co-rotation, and (c) {\it{Right:}} counter-rotation; obtained from the simulated data points constrained by observed distribution. The colorbar indicates the values of Galactic latitude $b$.}}
    \label{fig:mix}
\end{figure*}

As the Galactocentric distance to a source can be written as $R_{G} = 
\sqrt{R_{p}^2 + z^2}$, for non-coplanar orbits of the GCs we write the 
approximate circular velocity as
\begin{equation}
v_{R} = \bigg(\frac{d\Phi}{dR_{G}} R_{G}\bigg)^{1/2}
\end{equation}
\begin{equation}
\frac{d\Phi}{dR_{G}} = \frac{d\Phi}{dR_p} \frac{R_p}{R_G} + \frac{d\Phi}{dz} \frac{z}{R_{G}} \,,
\end{equation}
where $R_{p} = R_{G} \sin\theta$, is the component of galactocentric distance 
in the plane of the disk, and $z$ is the vertical height to the source. $\Phi = 
\Phi(R,z,d\Phi/dR,d\Phi/dz)$, is the potential of the system. \textsc{GalPot} 
takes as input the values of $R_{p}$ and $z$, and produces output $v_{R}$. As 
described in Section \ref{sec:results}, this is then used to compute the 
expected $v_{\rm los}$ for a given distance and direction.

\section{Analysis and Results} \label{sec:results}

For this assumed mass model, as described in Sec.~\ref{subsec:grav_pot}, we next 
find out the expected $l$ vs. $v_{\rm los}$ distribution numerically by transforming the velocity in Galactocentric frame to that in the observer frame. For this, 
we consider GCs as test particles, their angular positions ($\theta$, $\phi$) 
distributed uniformly on the surface of a sphere, and the Galactocentric 
distances $R_{G}$ having the same distribution as the observed one. For better 
statistics, we have used 200 times the number of data points in each $\Delta 
R_{G}=2$ kpc bin, so that the relative number of test particles in $l$ vs. 
$v_{\rm los}$ distribution represents the probablity density of observing a GC 
at that ($l$, $v_{\rm los}$). The angle $\psi$, between the rotation axis of 
the disk and the unit vector perpendicular to the GC's orbital plane, is taken 
as a random variable. From the circular velocity $v_{R}$ obtained using 
\textsc{GalPot}, we then compute $l$, $b$ and $v_{\rm los}$ for the test 
particles. Note that the value of $n_z=\cos\psi$ determines whether the GC is 
co-rotating ($\cos\psi>0$), counter-rotating ($\cos\psi<0$) or mixed (no 
constraint on $\psi$). These simulated $l$ vs. $v_{\rm los}$ distributions are 
shown in Fig.~\ref{fig:mix} left, middle and right panel for mixed, co-rotating 
and counter-rotating scenario, respectively. The color bar in the figures 
indicate the values of Galactic latitude $b$ in these plots. As expected, these 
distributions quantitatively deviate significantly from that of the neutral 
hydrogen and CO in the Galactic disk \citep{Kalberla08, Dame11} due to 
non-coplanar distribution of the GCs. 

\renewcommand{\arraystretch}{1.5}
\begin{table}[t]
	\centering
	\caption{2D K-S test for data vs. model}
	\label{tab:ks_test}	
	\begin{tabular}{lcc}
		\hline
		Dynamics & KS-statistic $d$ & $p$ value\\
		\hline
		Mixed- rotation & 0.119441 & 0.100220\\
		Co- rotation & 0.112045 & 0.148733\\
		Counter- rotation & 0.221819 & 0.000062\\
		\hline
	\end{tabular}
\end{table}

\begin{figure}
\centering
\includegraphics[width=0.49\textwidth]{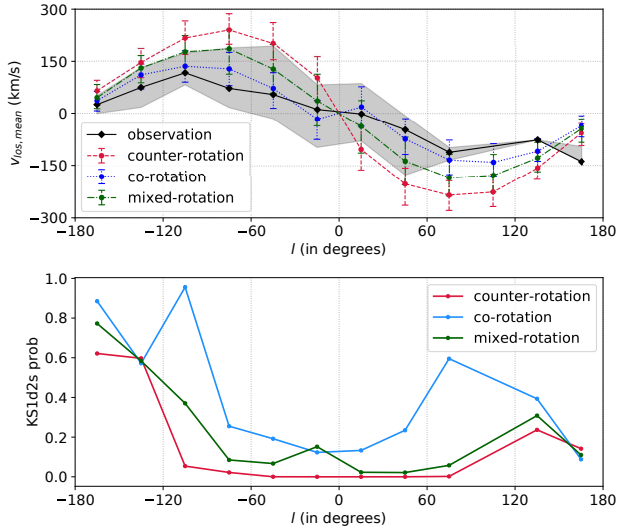}
\caption{\small{The top panel shows the median value of $v_{\rm los}$ in each bin with error bars representing the velocity range covered by first and third quartile values. The bottom panel shows the $p$ value for 1D K-S test between observation and models, in each $l$ bin with $\Delta l = 30^\circ$.}}
\label{fig:ks1d}
\end{figure}

We bin the $v_{\rm los}$ data into $30^\circ$ intervals over $l$, and find the 
median value of $v_{\rm los}$ in each $l-$bin for the various rotation models, 
as well as, the observed data. We show this in the upper panel of 
Fig.~\ref{fig:ks1d} along with error bars that represent the velocity range 
covered by first and third quartile values and thus encompasses 50\% of the 
data points. For the observed values, the shaded region indicates the region 
about the median for extrapolated values of the first and third quartiles. In 
the bin for $l$ range between $90^\circ-120^\circ$, no data for $v_{\rm los}$ 
is present from observation. It can be seen that the median values in each bin 
for co-rotation model are closer to that for observed data. To compare the 
observed distribution with the expected distributions from these three models, 
we used the Kolmogorov-Smirnov test \citep{Fasano87}. The test returns the K-S 
test statistic $d$ and the significance level $p$. Smaller $p$ values indicate 
that the data is significantly different from the model. The one-dimentional 
K-S statistics for the $v_{\rm los}$ distributions, considering the entire $l$ 
range, suggest that the observed distribution is more likely to match the 
co-rotation of GCs than counter or mixed rotation. We also perform a 1D K-S 
test for the $v_{\rm los}$ distribution in each $l-$bin of $30^\circ$ interval. 
The results are shown in the lower panel of Fig.~\ref{fig:ks1d}. The obtained 
$p$ values from 1D K-S test is higher for co-rotational model than counter or 
mixed rotation.

The K-S test is also well-suited to compare two samples of two-dimensional 
distributions obtained from the data and model. Here we consider the 2D dataset 
corresponding to $l$, $v_{\rm los}$ values. The obtained $d$ and $p$ values for 
comparison between the various modes of rotation with the observed data is 
shown in Table~\ref{tab:ks_test}. We find that the $p$ value is highest for 
co-rotation model. Thus, the null hypothesis that the two samples are drawn 
from the same distribution cannot be rejected for co-rotation model. However, 
the significance level being lower, the null hypothesis can be rejected for 
counter or mixed rotation. These results are consistent with that from 1D K-S 
tests. This clearly implies that, for the assumed mass model, the observed 
phase space distribution ($l$ - $v_{\rm los}$) of the GCs is consistent with a 
sample preferentially co-rotating with the Galactic disk.

We repeat our calculation of KS-statistic \textit{d} and the significance level 
\textit{p} after excluding the GCs from the dataset that may have originated 
from merging dwarf galaxies. Even after removing the GCs associated to Gaia 
Enceladus \citep{Myeong19}, the progenitor galaxy of the `sausage', and GCs 
from less massive progenitors like Sagittarius, Canis Major and Kraken 
\citep{Kruijssen18}, the co-rotation model is still found to be the preferred 
model in explaining the observed $l$ vs. $v_{\rm los}$ distribution. In fact, 
removing only the `Sausage' population yields $p = 0.49994$ for co-rotation, 
the highest amongst all the scenarios considered. On further excluding the 
other GCs from less massive parent dwarfs (Sagittarius, Canis Major, Kraken), 
the $p$ value is found to be 0.255308. This resonates with the observation that 
the Sequoia stars exhibit a strong retrograde motion, whereas the Sausage stars 
have no net rotation and move on predominantly radial orbits \citep{Myeong19}.

\section{Discussions} \label{sec:disc}

We use a standard Galactic mass distribution model from \textsc{GalPot} to understand the phase space distribution of the GCs of Milky Way Galaxy. For modeling, we restrict ourselves to simplified circular orbits of GCs and an axisymmetric gravitational potential of the  Milky Way. Note that the eccentricity of the GC orbits cannot be constrained at present from available observations alone in a model-independent manner. The uncertainties of the observed parameters depend on the inherent assumptions in modeling the underlying potential \citep{Simpson19}. With better data, when the eccentricity distribution is more constrained, this analysis can be further improved. Here, instead of considering a non-circular orbit of individual GCs, we have done an order of magnitude consistency check using the distribution of observed proper motion. The observed trend is found to be in broad agreement with the $Gaia$ measurements \citep{Eadie18,Vasiliev18}. Currently, reliable proper motion data is available for only 34 GCs from Gaia \citep{Watkins_2019}. We plan to carry out an extended but similar analysis with position ($l,v$), line-of-sight velocity and proper motion of the entire sample (expected to be soon available for the full sample) in the near future. The mass estimate of MW found in these studies and also in \cite{Watkins12, Posti19} are all consistent with each other within a factor of two. They have used a potential model which is similar to that used in the best-fit potential model of \textsc{GalPot}.

Please note, our analysis is a simple but complementary method to Jeans analysis of radial velocities of kinematic tracers like stars or star clusters to model the gravitational field of galaxies. However, the Jeans analysis requires measurement of the radial velocity, which is difficult, and it's dispersion, which, in turn, depends on the functional form of the circular velocity of the underlying potential. Also, the details of the orbits being unknown, these measurements have uncertainties from velocity anisotropy, stellar halo density profile at large distances \citep{Battaglia05, Bilek19}, etc. The radial distribution can be extrapolated for an incomplete GC survey, but the radial velocity dispersion, which is not a directly observable quantity, has to be deduced from the line-of-sight velocity measurements \citep{Binney82}, and will suffer from same uncertainties as in our analysis. It is worth mentioning that the conclusion drawn here is based on a static potential. Indeed, this is a simplification for the purpose of this study. The mass of the different components of the Galaxy changes through the significant amount of merging during the Galaxy evolution over Gyr timescale. A complete analysis including the full orbital evolution of the Galactic globular clusters is,  unfortunately, beyond the scope of the current analysis. However, based on the current observations, we expect that the main result of this analysis, that the GCs are preferentially corotating, will not significantly alter even when the time variation is included in the modeling.

The recent discovery of a tidal debris from what appears to be an ancient major merger event $\sim$10 Gyr ago \citep{Helmi18, Myeong18}, referred to as `Gaia sausage', is predicted to dominate the Galactic stellar halo at distances ranging from the MW bulge region to the MW halo's break radius at around 20-30 kpc \citep{Simion18, Deason18, Vincenzo19, Lancaster19}. \cite{Belokurov18} have estimated the virial mass of Gaia Enceladus, the progenitor galaxy, to be $M_{\rm vir}>10^{10}M_\odot$. Other such studies in the past have provided evidence of similar such accretion episodes like Sequoia event and merger of less massive progenitors like Canis Major, Kraken, and Sagittarius \citep{Ibata95, de_Boer15, Kruijssen18, Myeong19, Barba19}. Many of the GCs in our sample are associated with these merger events. Hence, we repeat the analysis by excluding those GCs that are associated with earlier merger events, to check whether preferentially co-rotating GC population is mostly due to the mergers. However, our analysis, after excluding these GCs, still shows (somewhat improved) accordance with the co-rotation model with the observed dynamics. This is indicative that the preferentially co-rotating model is not entirely due to the known accretion events.

Finally, a similar analysis can also be done with the satellite galaxies. A similar preliminary analysis shows marginally better match with the co-rotation model; however, the data is sparse. Out of 59 satellite galaxies of the Milky Way with known distance (within 0.5 Mpc), velocity information is available for only 28 \citep{Drlica15, Bechtol15, Koposov15}. Only recently, the data for satellite galaxies and dwarf spheroidals are reaching unprecedented refinement in the era of ongoing observational surveys \citep[e.g.,][]{Abbott18}. We plan to make use of the improved data set, including the recent proper motion measurements, with the complete sample of GCs and SGs for a detailed study in the future.

\section{Conclusions} \label{sec:conclu}

In this study, we have compared the observed ($l$, $v_{\rm los}$) phase space distribution of the Milky Way globular clusters with a simple scenario based on the standard mass model of the Galaxy. We use the best-fit potential model in \textsc{GalPot} for this, and compare the direct observables, position and line-of-sight velocity, to check if the GC dynamics is consistent with the adopted mass model. Multiple statistical measures show that the model with a co-rotating GC population is favoured over a counter-rotating or randomly rotating sample of GCs. We also find that even when the GCs associated with various progenitors of early merger events are excluded from the dataset, co-rotation is still found to be the preferred model. The recent compelling evidence of major merger events, along with the identification of GCs associated with these events, has significantly changed our perception of MW halo formation. The signatures of massive impacts during the evolutionary stage of Galaxy formation are retained in the substructures through their kinematical and chemical composition data. Extending such analysis with precise data of GC dynamics, including the reliable proper motion of complete sample, may be able to explain the overall co-rotation of the GC population fully and also put better constraints on the Milky Way mass model.

\section*{Acknowledgement}

SD acknowledges support from the Centre for Theoretical Studies, Indian 
Institute of Technology - Kharagpur, where part of this work was carried out. 
NR acknowledges support from the Infosys Foundation through the Infosys Young 
Investigator grant. This research has made use of the NASA/IPAC Extragalactic 
Database (NED), which is operated by the Jet Propulsion Laboratory, California 
Institute of Technology, under contract with the National Aeronautics and Space 
Administration.

\bibliographystyle{raa}
\bibliography{ms23} 
\end{document}